\documentclass[letterpaper,11pt]{article}

\usepackage[english]{babel}
\usepackage[utf8]{inputenc}
\usepackage{amsmath}
\usepackage{amssymb}
\usepackage{amsthm}
\usepackage{enumerate}
\usepackage{layout}
\usepackage{graphicx}
\usepackage{hyperref}
\usepackage{fancyhdr}
\usepackage{array}
\usepackage{color}
\usepackage{caption}
\usepackage{float}
\usepackage[margin=1in]{geometry}
\usepackage{wrapfig}

\newtheorem{theorem}{Theorem}[section]
\newtheorem{claim}[theorem]{Claim}
\newtheorem{corollary}[theorem]{Corollary}
\newtheorem{lemma}[theorem]{Lemma}
\newtheorem{defn}[theorem]{Definition}
\newcommand{\Alpha}{A}

\let\realbfseries=\bfseries
\def\bfseries{\realbfseries\boldmath}

\let\epsilon=\varepsilon

\newcommand\PARTITION{\textsc{Partition}}

\newcommand\backref[1]{#1}

\title{Cookie Clicker}

\author{%
  Erik D. Demaine%
    \thanks{CSAIL, Massachusetts Institute of Technology}
\and
  Hiro Ito%
    \thanks{University of Electro-Communications}
\and
  Stefan Langerman%
    \thanks{Directeur de Recherches du F.R.S.-FNRS,
      Universit\'e Libre de Bruxelles}
\and
  Jayson Lynch\footnotemark[1]
\and
  Mikhail Rudoy%
    \thanks{CSAIL, Massachusetts Institute of Technology; now at Google}
\and
  Kai Xiao\footnotemark[1]
}

\date{}

\begin{document}
\maketitle

\vspace{-2ex}

\begin{abstract}
Cookie Clicker\footnote{\url{http://orteil.dashnet.org/cookieclicker/}} is a popular online incremental game where the goal of the game is to generate as many cookies as possible. In the game you start with an initial cookie generation rate, and you can use cookies as currency to purchase various items that increase your cookie generation rate. In this paper, we analyze strategies for playing Cookie Clicker optimally. While simple to state, the game gives rise to interesting analysis involving ideas from NP-hardness, approximation algorithms, and dynamic programming.
\end{abstract}

\vspace{-2ex}

\tableofcontents

\newpage

\section{Introduction}

\begin{wrapfigure}{r}{3.25in}
  \includegraphics[width=\linewidth]{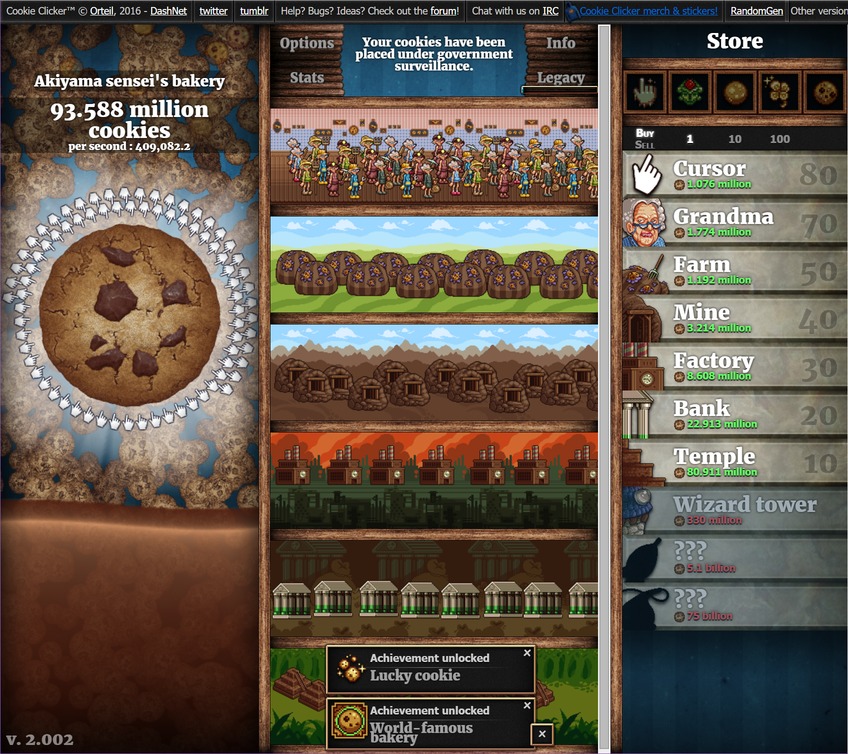}
  \caption{Screenshot of Cookie Clicker v.2.002.}
\end{wrapfigure}

In Cookie Clicker, your goal is to generate as many cookies as possible. There are two ways to generate cookies in the game -- you can click on a big cookie icon to bake a cookie, and you can purchase items that automatically generate cookies for you over time. We simplify these two mechanics into a single \emph{cookie generation rate}, which is defined as the number of cookies we generate per second. We do so by modeling the first mechanic (the ability to click on the big cookie icon to bake cookies) as a fixed initial cookie generation rate.

You can use the cookies you have generated as currency to purchase various items that increase your cookie generation rate. Items can be purchased multiple times, but after each item purchase, the item's cost will increase at an exponential rate, given by $C_n = C_1\cdot \alpha^{n-1}$, where $C_1$ is the cost of the first item and $C_n$ is the cost of item $n$. In the actual game, $\alpha = 1.15$. The real game has no explicit end condition, but in this paper we define two possible end conditions: reaching a certain number $M \in (0, \infty)$ of cookies, or reaching a certain cookie generation rate $R \in (1, \infty)$.

Cookie Clicker falls into a broader class of popular online games called \emph{incremental games} or \emph{idle games} \cite{Wikipedia}, in which the primary mechanic of the game is acquiring income and spending that income on income generators in order to acquire even more income. Some other well-known games in this genre include Adventure Capitalist, Cow Clicker, Clicker Heros, Shark Souls, Kittens Game, Egg Inc., and Sandcastle Builder (based on the xkcd comic 1190, \emph{Time}). Our analysis of Cookie Clicker involves solving a scheduling-style optimization problem, similar to prior work on job-scheduling algorithms where job utilization costs are involved \cite{scheduling}.

\subsection{Models}
Formally, the Cookie Clicker problem is as follows: Given an initial number of cookies $z$, an initial generation rate $r$, and a set of items, find the optimal sequence and timing of item purchases that optimizes some objective. There are multiple possible objectives that we could want to optimize for, but we focus on the following two: \\
\indent \textbf{1. Reaching $M$ cookies in as little time as possible.} (``M version'')\\
\indent \textbf{2. Reaching a generation rate of $R$ in as little time as possible.} (``R/rate-goal version'')

In most of this paper, unless stated otherwise we assume that you start with $z=0$ cookies and that the initial cookie generation rate from clicking on the big cookie icon is $r=1$. We will describe each item by a tuple $(x, y, \alpha)$, where $x \in (0, \infty)$ denotes how much the item will increase your cookie generation rate, $y \in (0, \infty)$ denotes the initial cost of the item, and $\alpha \in [1, \infty)$ denotes the multiplicative increase in item cost after each purchase. The case where $\alpha = 1$ for every item is a special case called the \emph{fixed-cost} case, which we analyze in Sections \ref{Fixed-Cost Cookie Clicker for 2 Items}, \ref{Fixed-Cost Cookie Clicker for $k$ Items}. Finally, we consider time to be continuous for the majority of the paper, and we analyze the case where time proceeds in discrete timesteps in Section \ref{Discrete Cookie Clicker is Strongly NP-hard}.

We will usually begin each case by discussing the $M$ version of the problem and then explain how to extend our results to the $R$ version. A third natural objective is to maximize the number of cookies $M$ or the generation rate $R$ achieved given a total amount of time $T$, and it can be solved by any algorithm that solves the first two variants using binary search on the values of $M$ and $R$.

\subsection{Results}

Our analysis of various versions of Cookie Clicker gives rise to interesting and varied results; refer to Table~\ref{table}. First, we present some general results, such as the fact that the optimal strategy involves a Buying Phase where items are purchased in some sequence as quickly as possible, and then a Waiting Phase where no items are purchased.

We begin our version-by-version analysis by examining the case where exactly 1 item is available for purchase, and we present formulas describing how many copies of the item should be purchased in both the fixed-cost case and increasing-cost case.

Next, we analyze cases involving 2 items. In the 2-item fixed-cost case, we prove that the optimal solution always involves consecutively buying some number of copies of one item, followed by consecutively buying some number of copies of the other item. 

Then, we analyze the case involving $k$ items. In the $k$-item fixed-cost case, a weakly polynomial time dynamic programming solution can be used to find the optimal sequence of items to buy, and in the increasing-cost case, a strongly polynomial time dynamic programming solution can be used. Additionally, a greedy algorithm can be devised with an approximation ratio that approaches $1$ for sufficiently large values of $M$.

Afterwards, we present negative results, including proofs of weak NP-hardness of the decision version of the problem of reaching a generation rate of $R$ as quickly as possible, as well as for a version of Cookie Clicker that allows you to start with a nonzero number of cookies. Finally, we define a discretized version of Cookie Clicker where decisions regarding whether or not to buy an item happen in discrete timesteps and prove strong NP-hardness for that version.

Python implementations of the dynamic programming solution and the greedy solutions to the General Cookie Clicker problem, and the dynamic programming solution to the Fixed-Cost Cookie Clicker problem, are available.%
\footnote{\url{https://github.com/kaixiao/Cookie-Clicker}}

\begin{table}
\centering
\begin{tabular}{ | m{13em} | m{5.2cm}| m{5.2cm} | } 
\hline
\textbf{Problem Variant} \newline & \textbf{Result for $M$ version} \newline & \textbf{Result for $R$ version} \newline \\ 
\hline
1-Item Fixed-Cost with item $(x, y, 1)$ [\S\ref{1-Item Cookie Clicker Solution}] & OPT takes $\approx \frac{y}{x}\ln \frac{M}{y}$ time \newline $O(1)$ to compute OPT 
& OPT takes $\approx \frac{y}{x}\ln \frac{R}{x}$ time \newline $O(1)$ to compute OPT\\ 
\hline
1-Item Increasing-Cost with item $(x, y, \alpha)$ [\S\ref{1-Item Cookie Clicker Solution}] & OPT will stop Buying Phase after $\log_{\alpha}\frac{M}{y}$ items \newline $O(1)$ to compute OPT
& OPT will stop Buying Phase after $\frac{R}{x}$ items \newline $O(1)$ to compute OPT\\ 
\hline
2-Item Fixed-Cost with items $(x_i, y_i, 1)$ where $y_2 > y_1$ [\S\ref{Fixed-Cost Cookie Clicker for 2 Items}] & OPT is of the form $[1, 1, \dots, 1, 2, \dots, 2]$ for large enough $M$ \newline $u_1\log_{\phi} u_2 + O(u_1)$ to compute OPT, where $u_i \approx \frac{y_i}{x_i}\log \frac{M}{y_i}$
& OPT is of the form $[1, 1, \dots, 1, 2, \dots, 2, 1, 1]$ for a small number of $1$'s at the end for large enough $R$. \\ 
\hline
$k$-Item Fixed-Cost with items $(x_i, y_i, 1)$ [\S\ref{Fixed-Cost Cookie Clicker for $k$ Items}] & $O(\max_i (\frac{Mx_ik}{y_i}))$ to compute OPT using Dynamic Programming  
& $O(kR)$ to compute OPT using Dynamic Programming \\ 
\hline
$k$-Item Increasing-Cost with items $(x_i, y_i, \alpha_i)$ [\S\ref{Increasing-Cost Cookie Clicker for $k$ Items}] & $O(\max_i(k\log_{\alpha_i}^k\frac{M}{y_i}))$ to compute OPT using Dynamic Programming \newline Greedy Algorithm has Approximation Ratio of $1+O(\frac{1}{\log M})$
& $O(\max_i (k(\frac{R}{x_i})^k))$ to compute OPT using Dynamic Programming \newline Weakly NP-hard by reduction from \PARTITION\\ 
\hline
$k$-Item Increasing-Cost with items $(x_i, y_i, \alpha_i)$ with Initial Cookies [\S\ref{Cookie Clicker with Initial Cookies}] & Weakly NP-hard by reduction from \PARTITION & Weakly NP-hard by reduction from $M$ version \\ 
\hline
$k$-Item Increasing-Cost with items $(x_i, y_i, \alpha_i)$ and Discrete Timesteps [\S\ref{Discrete Cookie Clicker is Strongly NP-hard}] & Strongly NP-hard by reduction from 3-\PARTITION & Strongly NP-hard by reduction from $M$ version \\
\hline
\end{tabular}
\caption{Summary of results.  Positive results are listed first, followed by negative results. OPT in the table denotes the optimal solution, and runtimes listed correspond to how long it takes to determine OPT.}
\label{table}
\end{table}

\subsection{Useful Tools}

Before proceeding to our main results, we develop some useful tools for finding optimal solutions for playing Cookie Clicker. We present these tools and show how they are applied to the 1-Item Case, but these tools are applicable to all versions of the game.

First, we can think of a game state as a tuple $(c, n_1, n_2, \dots, n_k)$ where $c$ is the number of cookies you have and $n_i$ is the quantity of item $i$ that you have. In the 1-Item case, the tuple is just $(c, n_1)$. Note that your current state in the game is entirely described by this tuple, and you can compute relevant quantities such as the cookie generation rate from this state.

In general, the following claim is true.

\begin{claim}\label{1}
If the next step of the optimal strategy involves buying an item at some point in the future, you should buy the item as soon as you can afford it.
\end{claim}
\begin{proof}
Suppose that from a given game state, a strategy involves buying the $i$th item $t$ seconds after you can afford it. Let $G$ denote the cookie generation rate at the game state, let $x$ denote the cookie generation rate increase from buying the item, and $y'$ denote the item's current cost. The net change in game state after these $t$ seconds is a gain of $1$ copy of item $i$ and a change in the amount of cookies by $Gt-y'$.

Then, consider the strategy that buys the $i$th item as soon as you can afford it and waits $t$ seconds afterwards. In this case, the net change in game state after these $t$ seconds is a gain of $1$ copy of item $i$ and a change in the amount of cookies by $(G+x)t-y'$. Thus, this new strategy results in the exact same result as the original strategy, except that it gains an extra $tx$ cookies, which is strictly better. Thus, an optimal strategy that intends to buy an item as its next step must buy it as soon as it can be afforded.
\end{proof}

This claim tells us that the optimal strategy will always wait until it can purchase an item and purchase it immediately, or it will wait until the target number of cookies $M$ is reached. Thus, the problem boils down to jumping between game states in which you have $0$ cookies and need to decide between waiting to reach $M$ cookies or purchasing an item. This means that the only thing we need to keep track of to determine our game state is $(n_1, n_2, \dots, n_k)$, the number of each item we have purchased. For example, in the 1-Item case, we only need to keep track of a single number $n_1$. Thus, we can define $t(n)$ to be the minimum amount of time needed to reach $M$ cookies from the game state $(0, n)$.

From Claim~\ref{1}, we can deduce that the optimal solution will have 2 phases. We will call them the \textbf{Buying Phase}, where the solution tries to buy items, and the \textbf{Waiting Phase}, where the solutions has bought all the items it needs and just waits until the items generate $M$ cookies. Every optimal solution can be represented by the sequence of items that should be bought in the Buying Phase.

\medskip

Next, we define some general notation that will be useful in the future.
\begin{defn}
$B([i_1, i_2, \dots, i_n], G, X, Y, \Alpha)$ is the amount of time needed to buy the items $i_1$, $i_2$, $\dots$, $i_n$ in order from an initial state with $0$ cookies, cookie generation rate $G$, and rate gains, initial costs, and cost increases described by the vectors $X = (x_1, x_2, \dots, x_k)$, $Y = (y_1, y_2, \dots, y_k)$, and $\Alpha = (\alpha_1, \alpha_2, \dots, \alpha_k)$ respectively.
\end{defn}
In cases where $\Alpha = (1, 1, \dots, 1)$, we may leave out the $\Alpha$ parameter in the notation. Sometimes, we also leave out $G$ if it is clear what $G$ is being referred to.

It is often helpful to bound the value of $B$. The following two results are true in general.
\begin{lemma}\label{2}
The following inequalities hold:
\begin{align*}
B([i_1, i_2, \dots, i_n], G, X, Y) &\leq \left(\sum_{k=1}^{n}y_{i_k}\right) / G \\
B([i_1, i_2, \dots, i_n], G, X, Y) &> \left(\sum_{k=1}^{n}y_{i_k}\right) / \left(G + \sum_{k=1}^nx_{i_k}\right)
\end{align*}
\end{lemma}
\begin{proof}
Let $G_{i_k}$ be the generation rate just before purchasing item $i_k$. Note that for all $k$, $G \leq G_{i_k} < G + \sum_{k=1}^nx_{i_k}$. Additionally, we know that $B([i_1, i_2, \dots, i_n], G, X, Y) =   \sum_{k=1}^{n}\frac{y_{i_k}}{G_{i_k}}$. Then we have that
\begin{equation*}
\sum_{k=1}^{n}\frac{y_{i_k}}{G_{i_k}} \leq \sum_{k=1}^{n} \frac{y_{i_k}}{G} = \left(\sum_{k=1}^{n}y_{i_k}\right) / G
\end{equation*}
and
\begin{equation*}
\sum_{k=1}^{n}\frac{y_{i_k}}{G_{i_k}} > \sum_{k=1}^{n} \frac{y_{i_k}}{G + \sum_{k=1}^nx_{i_k}} = \left(\sum_{k=1}^{n}y_{i_k}\right) / \left(G + \sum_{k=1}^nx_{i_k}\right)
\end{equation*}
\end{proof}

Next, we determine conditions for when buying an item is optimal. In general, we can determine an upper bound on the generation rate $G$ beyond which it will not be worth it to purchase any more items.

\begin{lemma}\label{3}
If your current cookie generation rate is $G$ and the items have rate increases and costs $(x_i, y_i)$, you should stop buying items if and only if
\begin{equation*}
G > \max_i\Big({\frac{Mx_i}{y_i} -x_i}\Big)
\end{equation*}
\end{lemma}

To prove the result, we begin by proving an alternate form of the result for the 1-Item example.

\begin{lemma}\label{4}
If your current cookie generation rate is $G$, you should buy an item with cost $y$ and rate increase $x$ if and only if
\begin{equation}
\frac{M}{y} \geq 1 + \frac{G}{x}.
\end{equation}
\end{lemma}
\begin{proof}
Suppose we are at a state where we have purchased $n_1$ items. Then, the optimal decision is either to purchase another item or to enter the Waiting Phase.

In the first case, the time taken is equal to
\begin{equation*}
\frac{y}{G} + t(n_1+1) \leq \frac{y}{G} + \frac{M}{G+x},
\end{equation*}
because a valid (but possibly not optimal) strategy from the state $(0, n_1+1)$ is to wait.

In the second case, the time taken is equal to $\frac{M}{G}$.

Thus, if it is the case that
\begin{equation}
\frac{y}{G} + \frac{M}{G+x} \leq \frac{M}{G},
\end{equation}
then we should go with the first strategy and purchase an item. Rearranging $(2)$ gives $(1)$. This means that if $(1)$ is satisfied, purchasing the item is better.

Now we show that if $(1)$ is not satisfied, then waiting is better. Suppose that $(1)$ is not satisfied, so $\frac{M}{y} < 1 + \frac{G}{x}$. Written in the form of $(2)$, this inequality becomes $\frac{y}{G} + \frac{M}{G+x} > \frac{M}{G}$. Note that for any rate $G' > G$ and $y' \geq y$, the inequality $\frac{M}{y'} < 1 + \frac{G'}{x}$ still holds. Now, suppose that the optimal strategy from this point forward is to purchase $k$ items for some $k > 0$ and then wait. Let $y_i$ and $G_i$ denote the price and cookie generation rate after $i$ item purchases from this point forward, and note that $G_m = G_{m-1}+x$. Then the time taken to achieve this equals $B([\underbrace{1, \dots, 1}_{k}], G, X, Y) + \frac{M}{G_k}$. Note that for any $k > 0$
\begin{align*}
B([\underbrace{1, \dots, 1}_{k}], G, X, Y) + \frac{M}{G_k} &= \frac{y}{G} + \frac{y_1}{G_1} + \dots + \frac{y_{k-2}}{G_{k-2}} + \frac{y_{k-1}}{G_{k-1}} + \frac{M}{G_{k-1}+x} \\
&> \frac{y}{G} + \frac{y_1}{G_1}  + \dots + \frac{y_{k-2}}{G_{k-2}} + \frac{M}{G_{k-1}} \\
&= B([\underbrace{1, \dots, 1}_{k-1}], G, X, Y) + \frac{M}{G_{k-1}}
\end{align*}
Thus, we have that
\begin{align*}
B([\underbrace{1, \dots, 1}_{k}], G, X, Y) + \frac{M}{G_k} &> B([\underbrace{1, \dots, 1}_{k-1}], G, X, Y) + \frac{M}{G_{k-1}} \\
&> B([\underbrace{1, \dots, 1}_{k-2}], G, X, Y) + \frac{M}{G_{k-2}} \\
&\cdots \\
&> B([1], G, X, Y) + \frac{M}{G_1} \\
&= \frac{y}{G}+\frac{M}{G+x} \\
&> \frac{M}{G}
\end{align*}

Thus, if $(1)$ is not satisfied, then the optimal strategy is to wait. This completes the proof of the lemma.
\end{proof}

Isolating $G$ from \backref{Lemma~\ref{4}} and reversing the statement, the following corollary holds.
\begin{corollary}\label{5}
If your current cookie generation rate is $G$, you should stop buying items with cost $y$ and rate increase $x$ if and only if
\begin{equation*}
G > \frac{Mx}{y} -x.
\end{equation*}
\end{corollary}

Applying \backref{Corollary~\ref{5}} to every item in $k$-Item Cookie Clicker proves \backref{Lemma~\ref{3}}.

\section{Positive Results}

\subsection{1-Item Cookie Clicker Solution}
\label{1-Item Cookie Clicker Solution}
Armed with the tools we developed in the previous section, we solve the 1-Item Cookie Clicker problem. Based on the results of the previous section, the optimal strategy is to purchase $k$ items for some $k \geq 0$ as soon as each item becomes affordable and then wait until we reach $M$ cookies. The total time that this takes is
\begin{equation*}
B([\underbrace{1, \dots, 1}_{k}], 1, x, y, \alpha) + \frac{M}{1+kx}= \sum_{n=0}^{k-1} \frac{y\cdot\alpha^n}{1+nx} + \frac{M}{1+kx}.
\end{equation*}
By \backref{Lemma~\ref{4}}, we know that if our current cookie generation rate is $G'$ and the current cost of the item is $y'$, then we should stop buying the item when $\frac{M}{y'} < 1 + \frac{G'}{x}$. After buying $k$ items, we have that $G' = 1+kx$ and $y' = y \cdot\alpha^k$.

In the special case of $\alpha = 1$, which we call the \textit{fixed-cost} case, the inequality becomes
\begin{equation*}
\frac{M}{y} < 1 + \frac{1+kx}{x} = 1+k+\frac{1}{x}
\end{equation*}
so $k$, the number of items we should buy before stopping, is the smallest integer larger than $\frac{M}{y}-1-\frac{1}{x}$. In this case, the total time the optimal solution takes is equal to
\begin{align*}
\sum_{n=0}^{k-1} \frac{y}{1+nx} + \frac{M}{1+kx} &= \frac{y}{x}\sum_{n=0}^{k-1} \frac{1}{1/x+n} + \frac{M}{1+kx} \\
&\approx \frac{y}{x}\sum_{n=0}^{k-1} \frac{1}{n} + \frac{M}{1+kx} \\
&\approx \frac{y}{x}\ln{k} + \frac{M}{1+kx} \\
&\approx \frac{y}{x}\ln{\frac{M}{y}} + \frac{M}{\frac{Mx}{y}} \\
&= \frac{y}{x}\left(\ln{\frac{M}{y}} + 1\right).
\end{align*}

If $\alpha > 1$, the inequality then becomes
\begin{align*}
\frac{M}{y\cdot\alpha^k} &< 1 + \frac{1+kx}{x} = 1+k+\frac{1}{x} \\
&\iff \frac{M}{y(1+k+\frac{1}{x})} < \alpha^k \\
&\iff \log_{\alpha}{\frac{M}{y}} - \log_{\alpha}{\left(1+k+\frac{1}{x}\right)} < k.
\end{align*}
In most reasonable cases, the $\log$ term on the left hand side of the inequality is fairly small, so $k \approx \log_{\alpha}{\frac{M}{y}}$.

Now, we derive similar results for the rate-goal version of the problem, where the goal is to reach a final rate of $R$ cookies. Note that in this version, there is no Waiting Phase, so only the Buying Phase needs to be analyzed. The optimal strategy for the rate-goal version is quite simple: buy the item whenever possible until the goal rate is reached. The goal rate of $R$ is reached after $k = \lceil\frac{R-1}{x}\rceil$ purchases of the item.

Then, for the fixed-cost case where $\alpha = 1$, the total time needed to reach the rate goal will be
\begin{align*}
\sum_{n=0}^{k-1} \frac{y}{1+nx} &= \frac{y}{x}\sum_{n=0}^{k-1} \frac{1}{1/x+n}\\
&\approx \frac{y}{x}\sum_{n=0}^{k-1} \frac{1}{n}\\
&\approx \frac{y}{x}\ln{k}\\
&= \frac{y}{x}\ln{\Big\lceil\frac{R-1}{x}\Big\rceil}.
\end{align*}
And for the increasing-cost case, the total time needed to reach the rate goal is
\begin{equation*}
\sum_{n=0}^{k-1} \frac{y\cdot\alpha^n}{1+nx}.
\end{equation*}

\subsection{Fixed-Cost Cookie Clicker for 2 Items}
\label{Fixed-Cost Cookie Clicker for 2 Items}

In this section and the next, we analyze the case where all the $\alpha$'s are equal to 1, which we call \textit{Fixed-Cost Cookie Clicker}. This is a natural starting point, as it corresponds to the economic situation in which items are fixed in price due to enough supply existing. You can think of this game as modeling the problem of optimizing discrete investments.

In the 2 Item Cookie Clicker problem, our goal is to reach $M$ cookies as quickly as we can, and the 2 items available are described by the tuples $(x_1, y_1)$ and $(x_2, y_2)$. These are defined analogously to the 1 item case. Without loss of generality, we can assume that $y_2 > y_1$. In this problem, we will also make the assumption that $\frac{x_2}{y_2} > \frac{x_1}{y_1}$. This is because if the reverse inequality held, then buying $\frac{y_2}{y_1}$ copies of item 1 gives a higher rate increase than buying a single instance of item 2, which means that it will never be optimal to buy item 2 if $M$ is large enough\footnote{$M$ must be large enough so that the effect of $\frac{y_2}{y_1}$ not being an integer is irrelevant in the long run}.

We can apply the general claims from the 1 item analysis here, so we know from \backref{Claim~\ref{1}} that the optimal strategy will jump between states where we have $0$ cookies, and that there is a Buying Phase and a Waiting Phase. As before, we can represent every optimal solution by the sequence of items that should be bought in the Buying Phase.

We now solve this problem. We will show that the sequence of items in the Buying Phase must be $[1, 1, \dots, 1, 2, 2, \dots, 2]$ when $M$ is large enough. Then, finding the optimal solution simply involves figuring out when to stop buying item 1 and when to start buying item 2, which can be determined in polynomial time.

To help us solve our problem, we will define the following.

\begin{defn}
The efficiency score of an item of cost $y$ and rate increase $x$ when you have generation rate $G$ is $\frac{y}{x}+\frac{y}{G}$.
\end{defn}

\begin{lemma}\label{6}
If you plan to buy both items consecutively, you should always buy the item with the lower efficiency score. In particular, let $T = (y_2-y_1) / (\frac{y_1}{x_1}-\frac{y_2}{x_2})$. Then, if $G < T$, you should purchase item $1$ followed by item $2$, and if $G > T$, you should purchase item $2$ followed by item $1$.
\end{lemma}

\begin{proof}
The efficiency score of an item dictates whether buying item $1$ then item $2$ is better than buying item $2$ then item $1$.

Suppose we have generation rate $G$. Then the cost of buying item $1$ then item $2$ is equal to $\frac{y_1}{G} + \frac{y_2}{G+x_1}$ and the cost of buying item $2$ then item $1$ is equal to $\frac{y_2}{G} + \frac{y_1}{G+x_2}$. If $G < T$, we can rearrange the inequality to get that
\begin{align*}
G &< \frac{y_2-y_1}{(\frac{y_1}{x_1}-\frac{y_2}{x_2})} \\
\iff \frac{y_1}{G} + \frac{y_1}{x_1} &< \frac{y_2}{G} + \frac{y_2}{x_2} \\
\iff y_1\Big(\frac{G+x_1}{x_1}\Big) &< y_2\Big(\frac{G+x_2}{x_2}\Big) \\
\iff y_1\Big(\frac{x_2}{G(G+x_2)}\Big) &< y_2\Big(\frac{x_1}{G(G+x_1)}\Big) \\
\iff y_1\Big(\frac{1}{G}-\frac{1}{G+x_2}\Big) &< y_2\Big(\frac{1}{G}-\frac{1}{G+x_1}\Big) \\
\iff \frac{y_1}{G} + \frac{y_2}{G+x_1} &< \frac{y_2}{G} + \frac{y_1}{G+x_2}.
\end{align*}

On the other hand, if $G > T$, then the reverse is true.
\end{proof}

Now, suppose that we have some optimal solution represented as a sequence of 1's and 2's. Now, we know that until the rate $G$ reaches $T$, we will never have a 2 followed by a 1. Similarly, after the rate $G$ passes $T$, we will never have a 1 followed by a 2. Thus, the final sequence must be of the following form.
\begin{equation*}
[1, 1, \dots 1, 1, 2, 2, \dots, 2, 2, 1, 1,  \dots, 1, 1].
\end{equation*}
Somewhere in the middle of the sequence of 2's, the generation rate reaches $T$.

Now, we will show that for large enough $M$, there will be no sequence of 1's at the end of the optimal solution.

\begin{theorem}\label{7}
Let $f(x_1, x_2, y_1, y_2) = \max\Big(2, \frac{2}{x_1} \cdot \frac{y_1+y_2}{\frac{y_1}{x_1}-\frac{y_2}{x_2}}\Big) $ If $M \geq (f(x_1, x_2, y_1, y_2) + 2) \cdot y_1$, then the optimal solution will have no 1's at the end.
\end{theorem}

\begin{proof}
Suppose for the sake of contradiction that there are $k$ 1's at the end of the sequence representing the optimal solution for some $k > 0$. We will show that replacing the final 1 with a 2 results in a better solution, which disproves the optimality of the original solution.

Denote that the rate before purchasing the final 1 in the optimal solution as $R$.

The time it takes to buy the final 1 and then wait until the goal $M$ is reached is equal to $\frac{y_1}{R} + \frac{M}{R+x_1}$. The time it takes to buy a 2 instead of the final 1 and then wait until the goal $M$ is equal to $\frac{y_2}{R} + \frac{M}{R+x_2}$. We want to prove that
\begin{equation*}
\frac{y_2}{R} + \frac{M}{R+x_2} < \frac{y_1}{R} + \frac{M}{R+x_1}
\end{equation*}
or equivalently that
\begin{equation}
\frac{M}{R+x_2} - \frac{M}{R+x_1} < \frac{y_1}{R} - \frac{y_2}{R}.
\end{equation}

Now, we know from \backref{Lemma~\ref{4}} and the fact that the optimal solution bought the final 1 that
\begin{equation*}
1 + \frac{R}{x_1} \leq \frac{M}{y_1} \iff \frac{M}{R+x_1} \geq \frac{y_1}{x_1}.
\end{equation*} 
Similarly, because the optimal solution can not buy another 2 after the final 1, we know that 
\begin{equation*}
1+\frac{R+x_1}{x_2} > \frac{M}{y_2} \iff \frac{M}{R+x_1+x_2} < \frac{y_2}{x_2}.
\end{equation*}
Combining the above two equations, we end up with
\begin{align*}
\frac{M}{R+x_1+x_2} - \frac{M}{R+x_1} &< \frac{y_2}{x_2} - \frac{y_1}{x_1} \\
\iff \frac{M}{R+x_2} - \frac{M}{R+x_1} &< \frac{y_2}{x_2} - \frac{y_1}{x_1} + \frac{M}{R+x_2} - \frac{M}{R+x_1+x_2}  \\
&= \frac{y_2}{x_2} - \frac{y_1}{x_1} + \frac{Mx_1}{(R+x_2)(R+x_1+x_2)} \\
&< \frac{y_2}{x_2} - \frac{y_1}{x_1} + \frac{Mx_1}{R^2}.
\end{align*}

Thus, to prove (3), we just have to prove that
\begin{equation*}
\frac{y_2}{x_2} - \frac{y_1}{x_1} + \frac{Mx_1}{R^2} < \frac{y_1}{R} - \frac{y_2}{R}
\end{equation*}
or that
\begin{equation*}
\frac{Mx_1}{R^2} + \frac{y_2-y_1}{R} < \frac{y_1}{x_1} - \frac{y_2}{x_2}
\end{equation*}

Now note that because the optimal solution can not buy another 1 after the final 1,
\begin{equation*}
1+\frac{R+x_1}{x_1} > \frac{M}{y_1} \iff R > \Big(\frac{M}{y_1}-2\Big)x_1.
\end{equation*}
Because $M  \geq (f(x_1, x_2, y_1, y_2)+2) \cdot y_1$ and $f(x_1, x_2, y_1, y_2) \geq 2$, we can deduce that
\begin{align*}
M &\geq 4y_1\\
\iff M/2 &\geq 2y_1\\
\iff M-2y_1 &\geq M/2,
\end{align*}
and therefore
\begin{equation*}
R > \left( \frac{M}{y_1} -2 \right) x_1 = \frac{x_1}{y_1} (M - 2y_1)
\geq \frac{x_1}{y_1} \cdot \frac{M}{2} = \frac{Mx_1}{2y_1}.
\end{equation*}
Thus, $\frac{Mx_1}{R^2} =\frac{M}{R}\frac{x_1}{R} < \frac{2y_1}{x_1}\frac{x_1}{R} = \frac{2y_1}{R}$. Using this, all we have to prove now is that
\begin{equation*}
\frac{2y_1}{R} + \frac{y_2-y_1}{R}  = \frac{y_1+y_2}{R} < \frac{y_1}{x_1} - \frac{y_2}{x_2}
\end{equation*}
or equivalently that
\begin{equation*}
\frac{y_1+y_2}{\frac{y_1}{x_1}-\frac{y_2}{x_2}} < R.
\end{equation*}
But this is true because
\begin{equation*}
R > \frac{Mx_1}{2y_1} \geq \frac{(f(x_1, x_2, y_1, y_2)+2) \cdot x_1}{2} > \frac{f(x_1, x_2, y_1, y_2) \cdot x_1}{2} \geq \frac{y_1+y_2}{\frac{y_1}{x_1}-\frac{y_2}{x_2}} .
\end{equation*}
\end{proof}

\begin{figure}[h!]
  \centering
  \includegraphics[width=9.5cm]{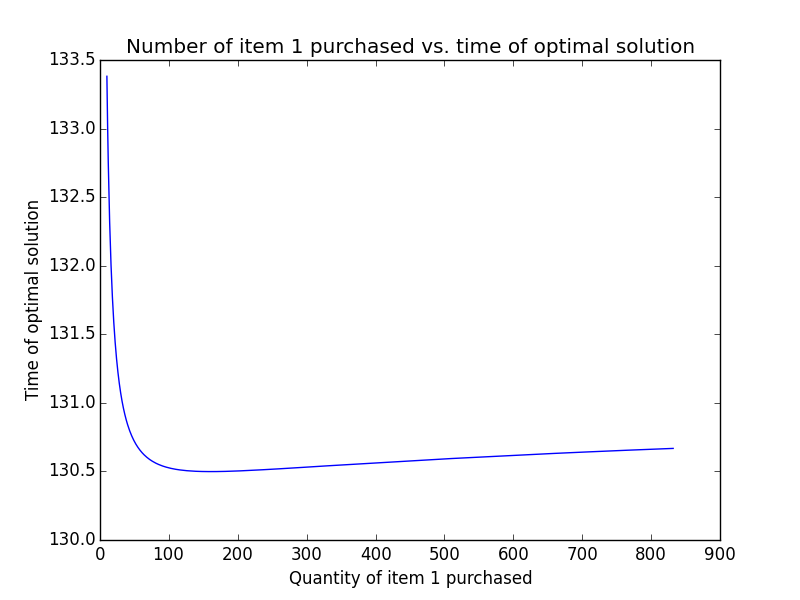}
  \caption{Here, $M = 60000$, $X = [10, 100]$, $Y= [72, 700]$, $T= 3140$, and $G = 1611$ at the minimum of the curve, which corresponds to the correct number of item 1's to purchase in the optimal solution. For most parameter settings, there is exactly one local minimum. However, this is not always the case.}
  \label{f1}
\end{figure}

Thus, we have shown that for large enough $M$, the optimal solution will be of the form
\begin{equation*}
[1, 1, \dots 1, 1, 2, 2, \dots, 2, 2],
\end{equation*}
where the 1's only appear if the total generation rate at that point is less than the threshold $T$. We can experimentally verify that the point at which the optimal solution transitions from 1's to 2's is not exactly $T$, but is usually close to $\frac{T}{2}$. An example of this is displayed in \backref{Figure~\ref{f1}}. For every integer $i$, we can consider the optimal strategy that starts off by buying exactly $i$ copies of item 1 then transitioning to item 2. \backref{Figure~\ref{f1}} plots the amount of time each optimal solution takes.

We then have the following corollary
\begin{corollary}
Fixed-Cost Cookie Clicker for 2 Items can be solved in $u_1 \log_\phi{u_2} + O(u_1)$ time, where $u_1 = O(\frac{y_1}{x_1}\log \frac{M}{y_1})$ and $u_2 = O(\frac{y_2}{x_2}\log \frac{M}{y_2})$.
\end{corollary}
\begin{proof}
To solve Fixed-Cost Cookie Clicker for 2 Items, we just have to find the optimal number of 1's to buy before transitioning to 2's and subsequently solving the 1-Item Cookie Clicker game. Finding this optimal number involves maximizing a function of two bounded discrete variables (the number of 1's to buy and the number of 2's to buy), which can be done in polynomial time.

In particular, let us define
\begin{equation*}
W(r,s) = B([\underbrace{1, \dots, 1}_{r}, \underbrace{2, \dots, 2}_{s}]).
\end{equation*}

Then the function $W$ is unimodal in $s$, because for a fixed $r$, we are essentially solving 1-Item Cookie Clicker using the item 2. We can obtain rough upper bounds for each of $r$ and $s$, which we denote by $u_1 = O(\frac{y_1}{x_1}\log \frac{M}{y_1})$ and $u_2 = O(\frac{y_2}{x_2}\log \frac{M}{y_2})$, by considering an instance of 1-Item Cookie Clicker with just item 1 or just item 2. Then, we can find an optimal solution in $u_1 \log_\phi{u_2} + O(u_1)$ \cite{UnimodalOptimization_ESA2005}.
\end{proof}

Next, we analyze the rate-goal version of Fixed-Cost Cookie Clicker for 2 Items and obtain a similar result about the structure of solutions.
\begin{theorem}
For large enough $R$, any optimal solution to the rate-goal version of Fixed-Cost Cookie Clicker for 2 Items must be of the form
\begin{equation*}
[1, 1, \dots 1, 1, 2, 2, \dots, 2, 2, 1, \dots, 1],
\end{equation*}
where the number of $1$'s at the end is upper bounded by $\lceil x_2/x_1 \rceil \Big(\Big\lfloor1/(\frac{x_2}{x_1}-\frac{y_2}{y_1})\Big\rfloor + 1\Big)$.
\end{theorem}

This theorem lets us restrict the space of possible solutions, and we can use brute force on the number of $1$'s at the end to obtain an algorithm that finds the optimal solution in the same amount of time as in the $M$ version, up to a multiplicative factor corresponding to the brute force search.

\begin{proof}
First, we know using the same swapping argument as before that the solution must be of the form
\begin{equation*}
[1, 1, \dots 1, 1, 2, 2, \dots, 2, 2, 1,1,  \dots, 1, 1].
\end{equation*}
Thus, we only have to show that the number of $1$'s at the end must be small. The primary idea is that if item 2 is indeed more efficient, you should be able to replace $\Big\lfloor \frac{x_2}{x_1} \Big\rfloor$ copies of item 1 with a single copy of item 2, giving a greater rate increase in less time, which can only improve your solution. This argument isn't always easy to prove, depending on the costs and rate increases of items 1 and 2. Thus, we will instead argue that buying $n$ copies of item $1$ at the end is worse than buying $m$ copies of item of item 2 for some positive integers $n$ and $m$. In particular, the minimal $n$ such that the rational number $\frac{n}{m}$ satisfies the equation
\begin{equation*}
\frac{x_2}{x_1} > \frac{n}{m} > \frac{y_2}{y_1}
\end{equation*}
is sufficient for large enough $R$. It is difficult to find the minimal $n$, but we know that if we choose an $m$ such that $\frac{1}{m} < \frac{x_2}{x_1} - \frac{y_2}{y_1}$, then there must exist some $n$ that satisfies the inequality and is less than $\lceil x_2/x_1 \rceil m$. Thus, one valid choice of $m$ is $\Big\lfloor1/(\frac{x_2}{x_1} - \frac{y_2}{y_1})\Big\rfloor+1$, and $n$ is then upper bounded by $\lceil x_2/x_1 \rceil \Big(\Big\lfloor1/(\frac{x_2}{x_1}-\frac{y_2}{y_1})\Big\rfloor + 1\Big)$. This $n$ upper bounds the number of trailing 1's we can have in any solution.

To prove the above claim, consider replacing $n$ trailing copies of item 1 in some solution $S$ with $m$ copies of item 2 to produce the new solution $S'$:
\begin{equation*}
[1, \dots 1, 2, \dots, 2, 1, \dots, 1, \underbrace{1, \dots, 1}_{n}] \rightarrow [1, \dots 1, 2, \dots, 2, 1, \dots, 1, \underbrace{2, \dots, 2}_{m}].
\end{equation*}

The rate increase from the $n$ trailing item 1's in $S$ is $nx_1$, and the rate increase from the trailing item 2's in $S'$ is $mx_2$. Based on our definition of $n$ and $m$, we know that the rate increase from the trailing item 2's in $S'$ is higher, so $S'$ will also reach the goal rate of $R$.

Now, we just have to show that the solution $S'$ takes less time than the solution $S$. Suppose that the generation rate before purchasing the $n$ copies of item 1 is $G$. The amount of time it takes to buy the $n$ copies of item 1 in $S$ is
\begin{equation*}
B([\underbrace{1, \dots, 1}_{n} ], G) > n\frac{y_1}{G+nx_1}.
\end{equation*}
The amount of time it takes to buy the $m$ copies of item 2 in $S'$ is
\begin{equation*}
B([\underbrace{2, \dots, 2}_{m} ], G)  \leq m\frac{y_2}{G},
\end{equation*}
where both inequalities are derived from \backref{Lemma~\ref{2}}.

Thus, we just have to show that
\begin{equation*}
n\frac{y_1}{G+nx_1} > m\frac{y_2}{G}
\end{equation*}
or equivalently, that
\begin{equation*}
ny_1G > my_2G + mnx_1y_2
\end{equation*}
or
\begin{equation}
G > \frac{mnx_1y_2}{ny_1-my_2}
\end{equation}
Let $R'$ be the fraction on the right hand side of (4). The denominator of $R'$ is positive from the definition of $n$ and $m$, so as long as $G$ is large enough, buying $m$ copies of item 2 at the end also takes less time than buying $n$ copies of item 1. Thus, as long as $R > R' + nx_1$, the optimal solution will have at most $n$ trailing 1's.
\end{proof}

\subsection{Fixed-Cost Cookie Clicker for $k$ Items}
\label{Fixed-Cost Cookie Clicker for $k$ Items}

A natural follow-up is to extend this problem from 2 items to $k$ items. Here, we present a weakly polynomial time dynamic programming solution and discuss an attempt using local optimizations to achieve a faster runtime.

\subsubsection{Dynamic Programming Solution}
\label{DP1}

In the fixed-cost case, the items do not change in price over time. Thus, an equivalent way to fully represent the game state in this case is to specify the current generation rate. Using \backref{Lemma~\ref{3}}, the final generation rate is bounded above by $\max_i\big(\frac{Mx_i}{y_i}\big)$. Let DP[$r$] denote the minimal amount of time needed to reach a goal of $M$ from a state where you have $0$ cookies and generation rate $r$. The dynamic program can be solved by the following recurrence:
\begin{equation*}
DP[r] = \min\Big(\frac{M}{r}, \min_i\Big({\frac{y_i}{r} + DP[r+x_i]}\Big)\Big),
\end{equation*}
where the $\frac{M}{r}$ term corresponds to entering the Waiting Phase after achieving a generation rate of $r$.
If there are $k$ items, then solving the original problem, which is equivalent to finding the value of DP[$1$], takes $O(\max_i (\frac{Mx_ik}{y_i}))$.

For the rate-goal version of this problem, we can define the same subproblem DP[$r$], and the recurrence can be modified slightly to
\[ DP[r] = 
 \begin{cases} 
       \min_i({\frac{y_i}{r} + DP[r+x_i]}) & \text{if\ } r < R, \\
       0 & \text{otherwise}.
   \end{cases}
\]
Again, the final problem reduces to finding DP[$1$], and the total runtime is $O(Rk)$.

\subsubsection{Local Optimizations}

One idea for a faster algorithm is to use local optimizations on a given solution sequence to try to obtain a better one.

As we saw from the analysis of Fixed-Cost Cookie Clicker for 2 Items, one example of a local optimization is a ``swap" that involves swapping consecutive elements in a solution if doing so improves the solution. Another natural local optimization, which we saw in the proof of \backref{Theorem~\ref{7}}, was the replacement of one item in a solution sequence with another.

Thus, to try to solve this problem, we tried using random local optimizations on a randomly generated initial solution until it reached a point where local optimizations could no longer improve the solution. The local optimizations we used included:
\begin{enumerate}
\item Adding an item to the solution at a specific index
\item Deleting an item from the solution at a specific index
\item Replacing one item with a different item at a specific index
\item Moving an item from one index to another. If an item is moved from index $i+1$ to $i$, this is equivalent to a ``swap".
\item Sorting the solution so that the cheapest items come first
\end{enumerate}
In general, these local optimizations would improve the initial solution sequences. In some cases these local optimizations would result in a final solution that matched the globally optimal solution computed using dynamic programming. However, in other cases, these local optimizations get stuck at a local optimum, which are points that are worse than the global optimum, but can not be improved any further using any of these local optimization operations. Thus, this leads us to the conclusion that our list of local optimizations is insufficient for finding a global solution, and any proof that relies solely on these operations will not be able to show global optimality of a solution.

\subsection{Increasing-Cost Cookie Clicker for $k$ Items}
\label{Increasing-Cost Cookie Clicker for $k$ Items}

From this point forward, we assume that the $\alpha$'s, the cost increase rates, all satisfy $\alpha > 1$. This is how the original Cookie Clicker game works, and is also a reasonable assumption. It corresponds to the economic situation in which items increase in price due to limited supply.
\subsubsection{Dynamic Programming Solution}
In this section, we present a dynamic programming solution which finds optimal solutions in $O(k\log^k_{\alpha}(\frac{M}{y}))$ time.

For simplicity, let us solve the problem for the case with $2$ items first. Recall that your state in the game is completely described by the tuple $(n_1, n_2)$, where $n_i$ denotes the quantity of item $i$ you have purchased. Note that it will never be worth it to buy an item if the item costs more than the goal $M$. This gives us upper bounds on $n_1$ and $n_2$, namely, $n_i < \log_{\alpha_i}{\frac{M}{y_i}}+1$. This motivates us to define
\begin{equation*}
DP[a][b] := \text{the minimum time it takes to reach $M$ from the state (a,b)}.
\end{equation*}
Let the cookie generation rate at state $(a,b)$ be represented by $g_{ab} = 1+ax_1+bx_2$. From the state $(a,b)$, the optimal strategy is one of the following three choices: entering the Waiting Phase, buying item 1, or buying item 2. We can then derive the recurrence
\begin{equation*}
DP[a][b] = \min\Big(\frac{M}{g_{ab}}, \frac{y_1\cdot\alpha_1^a}{g_{ab}} + DP[a+1, b], \frac{y_2\cdot\alpha_2^b}{g_{ab}} + DP[a, b+1]\Big)
\end{equation*}
corresponding to each of those three choices. If $A$ and $B$ are the upper bounds for $n_1$ and $n_2$ respectively, then we can revise the recursive formulas for $DP[A][b]$ and $DP[a][B]$ for any $a$ and $b$ in the proper range to only correspond to two choices (e.g. for $DP[A][b]$, you can either buy item 2 or wait). We only need to initialize the value $DP[A][B] = \frac{M}{g_{AB}}$ and then use the recurrence to fill out the rest of the dynamic programming table. Finally, our answer is $DP[0][0]$.

Initializing the single boundary value takes $O(1)$ time. Then, filling out the rest of the $A \text{x} B$ table takes $O(AB)$ time, so the total runtime is $O(AB)$.

This dynamic programming approach can easily be extended to the $k$-item problem. As before, one can note that the game state of the $k$-item problem is described entirely by the $k$-tuple $(n_1, n_2, \dots, n_k)$, where $n_i$ is the quantity of item $i$ that you have purchased. We can use the same upper bound $n_i < \log_{\alpha_i}(\frac{M}{y_i})+1$. Let $N_i=\log_{\alpha_i}(\frac{M}{y_i})+1$ denote the upper bounds for each $n_i$.

We can similarly define $DP[(n_1, n_2, \dots, n_k)]$ to be the minimum time it takes to reach M from the state $(n_1, n_2, \dots, n_k)$. Then, filling in any square in the grid involves checking the solutions the adjacent squares and doing an $O(1)$ computation for each adjacent square. In total, this takes $O(k)$ time. The only square we need to initialize is the corner $DP[(N_1, N_2, \dots, N_k)]$. Thus, the total time complexity of this program is $O(k\prod_{i=1}^{k}{N_i}) = O(kN_{max}^k) = \boxed{O\Big(k\log^k_{\alpha}\Big(\frac{M}{y}\Big)\Big)}$.

For the rate-goal version of the problem, we can also use dynamic programming to compute the optimal solution, but the complexity is significantly worse. We also define $DP[(n_1, n_2, \dots, n_k)]$ to be the minimum time it takes to reach $R$ from the state $(n_1, n_2, \dots, n_k)$, but we can only obtain a naive upper bound for each $n_i$ of $N_i = \frac{R}{x_i}$. This upper bound does not have a $\log$ like the upper bound for the $M$ version of the problem because we can not take advantage of the fact that costs increase. No matter how costly the items get, we still have to get to the final rate $R$. Using this upper bound, this dynamic programming approach takes $O(k\prod_{i=1}^{k}{N_i}) = O(kN_{max}^k) = \boxed{O\Big(k\Big(\frac{R}{x}\Big)^k\Big)}$.

\subsubsection{Greedy Solutions}

\paragraph{Natural Greedy Solution: }One greedy solution that arises naturally in normal gameplay involves buying the item that has the highest rate increase to cost ratio $\frac{x_i}{y_i}$. This is the calculation that most human players do when playing the game unaided. For the 2 item case, given most reasonable setting of the parameters, this approach actually performs fairly well. However, for certain settings of the parameters, this approach can be quite bad. For example, take $M = 10000, (x_1, y_1, \alpha_1) = (1, 10, 2), (x_2, y_2, \alpha_2) = (10000, 9999, 2)$. In this case, the second item has a much higher rate increase to cost ratio, which means that the natural greedy solution would save up for a long time to purchase the second item. However, it is much better to purchase the first item and increase your generation rate incrementally.

\paragraph{Efficiency Score Greedy Solution: }As we saw in the analysis of the fixed-cost case, the efficiency score was a helpful metric to determine which item to buy. Another idea for a greedy algorithm is to compute the efficiency score of each item and always choose the item with the lowest efficiency score. This approach is generally very close to optimal. In fact, we can prove approximation guarantees for this greedy solution.

\subsubsection {Approximation Ratio}
Once again, we will begin with the 2 Item case. We derive an approximation ratio for the \backref{Efficiency Score Greedy Solution} that approaches $1$ for sufficiently large $M$. Specifically, we will prove the following theorem.
\begin{theorem}
The \backref{Efficiency Score Greedy Solution} obtains an approximation ratio of $1 + O\Big(\frac{1}{\log M}\Big)$ for sufficiently large $M$.
\end{theorem}
\begin{proof}

The main idea is to use two propositions:

\begin{enumerate}
\item \textbf{Proposition 1}: Before you're anywhere close to reaching the goal $M$, you'll want to purchase at least one more copy of item 1 and at least one more copy of item 2
\item \textbf{Proposition 2}: When $G$ is ``big enough", $\alpha > 1$ means that the most efficient item is locally and globally optimal
\end{enumerate}
Using these two propositions, the greedy solution can be compared to the optimal solution in three phases.

\begin{itemize}
\item \textbf{Phase 1}: In this phase, the greedy algorithm reaches a certain generation rate $G$, which corresponds to the ``big enough" $G$ referred to in Proposition 2. ``Big enough" is quantified in Claim~\ref{8}. The time taken is labeled as $T_1$.
\item \textbf{Phase 2}: The optimal algorithm will buy some amount of each of the 2 items, reaching $c_1$ copies of item 1 and $c_2$ of item 2 (the game state $(c_1, c_2)$) near the tail end of the Buying Phase. From this point onwards, the optimal algorithm will only purchase one type of item --- this is Phase 3.

We will show that the greedy algorithm also reaches the state $(c_1, c_2)$. The interval between the end of Phase 1 and reaching the state $(c_1, c_2)$ is Phase 2. In this phase, the greedy algorithm and optimal algorithm would be equivalent from the same starting state\footnote{In the final analysis, we compare the time needed by the optimal algorithm starting from a worse state than the greedy one, so the greedy solution actually takes less time in Phase 2.}. If we let $T_2$ denote the amount of time the greedy algorithm uses in Phase 2, and we let $O_2$ be the amount of time optimal algorithm takes to reach $(c_1, c_2)$ from $(0, 0)$, then $T_2 < O_2$.
\item \textbf{Phase 3}: In this phase, Proposition 1 listed above is no longer true. Let $w$ denote the number of consecutive copies of a single item that the optimal algorithm buys at the very end of the Buying Phase. If $T_3$ is the amount of time the greedy algorithm takes in this phase and $O_3$ is the amount of time the optimal solution takes, we can show that $T_3 < O_3(\alpha_1 (w\cdot \frac{x_2}{x_1}+1))$ if the single item is item 2, and $T_3 < O_3(\alpha_2(w+\frac{x_2}{x_1}))$ if the single item is item 1. Furthermore, we can bound the value of $w$, so $T_3 < d_2O_3$ for a constant $d_2$ that depends only on the $x$'s and $\alpha$'s.
\end{itemize}

Here is an example of the greedy solution compared to the optimal solution, with the three phases labeled. Take $M = 100000$, $X = \{10, 90\}$, $Y = \{80, 800\}$,
$\alpha = \{1.2, 1.1\}$. \\ \\
Optimal Solution: $[\underbrace{1, 1, 1, 1, 1, 2, 2, 2, 2, 2, 2, 2, 2, 1, 2, 2, 1, 2, 2, 1, 2, 2, 1, 2, 2, 1, 2, 2, 1}_{O_2}, \underbrace{2}_{O_3}]$ \\
Greedy Solution: $[\underbrace{1, 1, 1, 1, 1, 1, 2, 2}_{T_1}, \underbrace{2, 2, 2, 2, 2, 2, 2, 2, 1, 2, 2, 1, 2, 2, 1, 2, 2, 1, 2, 2, 1}_{T_2}, \underbrace{2}_{T_3}]$

To begin, we will show that, when Proposition 1 is true and $G$ is sufficiently large, then Proposition 2 is true.

Suppose without loss of generality that we are currently at a state where $B([1,2]) < B([2,1])$. We will say that this state has the $E_1$ property, meaning that item $1$ is currently more efficient, and the locally optimal decision would be to buy item $1$.

Due to Proposition 1, we know that any optimal solution from this current state will purchase $k$ copies of item $2$ followed by a copy of item $1$ for some number $k$. We want to show that given a solution where $k > 0$, we can produce a better solution by choosing to purchase item $1$ first before purchasing any of the $k$ copies of item $2$. That is, given that $B([2,1]) > B([1,2])$, then
\begin{equation*}
B([2,2,\dots 2,1]) > B([1,2,2,\dots 2])
\end{equation*}
which can be argued by sequentially showing that
\begin{align*}
B([2,2,\dots 2, 2,1]) &> B([2,2\dots 2, 1, 2]) \\
\ \, &> B([2,2\dots 1, 2, 2])  \\
&\ \ \ \ \ \vdots \\
\ \, &> B([1,2,2,\dots 2, 2]).
\end{align*}

This would be true if the $E_1$ property still held after each purchase of item $2$. Intuitively, it should, because purchasing item $2$ actually makes future purchases of item $2$ more expensive. However, the $E_1$ property doesn't necessarily hold after some number of purchases of item $2$ because when the generation rate goes up, the more expensive item, which could be item 2, could become the more efficient item. Thus, we will rely on the following claim.

\begin{claim}\label{8}
Let $q_2 = \frac{G}{x_2}$. Suppose that a state with generation rate $G$ satisfies the $E_1$ property. Then the next item that should be purchased is item 1 if $q_2^2+2q_2 \geq \frac{1}{\alpha_2-1}$.
\end{claim}
\begin{proof}
Let $y_1$ and $y_2$ denote the current costs of item 1 and item 2 after factoring in the cost increases. If item 1 has a lower efficiency score than item 2 at generation rate $G$, then
\begin{equation*}
\frac{y_1}{x_1} + \frac{y_1}{G} \leq \frac{y_2}{x_2} + \frac{y_2}{G}.
\end{equation*}
First, we find conditions when $B([1, 2]) < B([2, 1]) \implies B([2, 1, 2]) < B([2, 2, 1])$.

To show that $B([2, 1, 2])  < B([2, 2, 1])$, we need to prove that
\begin{equation*}
\frac{y_1}{x_1} + \frac{y_1}{G+x_2} \leq \alpha_2\Big(\frac{y_2}{x_2} + \frac{y_2}{G+x_2}\Big).
\end{equation*}
We know that
\begin{equation*}
\frac{y_1}{x_1} + \frac{y_1}{G+x_2}  < \frac{y_1}{x_1} + \frac{y_1}{G} \leq \frac{y_2}{x_2} + \frac{y_2}{G},
\end{equation*}
so we just have to show that
\begin{equation*}
\frac{y_2}{x_2} + \frac{y_2}{G} \leq \alpha_2\Big(\frac{y_2}{x_2} + \frac{y_2}{G+x_2}\Big).
\end{equation*}
Notice that
\begin{align*}
    a&\frac{y_2}{x_2} + \frac{y_2}{G} &\leq \alpha_2\Big(\frac{y_2}{x_2} + \frac{y_2}{G+x_2}\Big) \\ 
\iff&& \frac{y_2}{G} &\leq (\alpha_2-1) \frac{y_2}{x_2} + \frac{\alpha_2y_2}{G+x_2} \\
\iff&& \frac{1}{x_2q_2} &\leq (\alpha_2-1) \frac{1}{x_2} + \frac{\alpha_2}{x_2q_2+x_2} \\
\iff&& q_2+1 &\leq (\alpha_2-1) q_2(q_2+1) + \alpha_2q_2 \\
\iff&& 1 &\leq (\alpha_2-1)(q_2^2+2q_2) \\
\iff&& \frac{1}{\alpha_2-1} &\leq q_2^2+2q_2.
\end{align*}

This is the original assumption in \backref{Claim~\ref{8}}. Thus, we know that $B([2, 1, 2])  < B([2, 2, 1])$ if the original condition holds. We also know that $B([1, 2, 2]) < B([2, 1, 2])$, so the optimal solution from our current state can not start with a $[2, 1]$ sequence or a $[2, 2, 1]$ sequence.

Now, we claim that this holds for any string of 2's in the beginning, that is, $B([2, 2, \dots2, 1, 2]) < B([2, 2, \dots 2, 2, 1])$. This is true because at the point where $[1, 2]$ needs to be compared to $[2, 1]$, the generation rate $G'$ satisfies $G' > G$, so $q_2' > q_2$. Then, the same argument holds because
\begin{equation*}
\frac{1}{\alpha_2-1} \leq q_2^2+2q_2 < q_2'^2+2q_2'.
\end{equation*}
Thus, if $G$ is large enough and we are in a state that satisfies the $E_1$ property, then the next item that should be purchased must be item 1.
\end{proof}

Next, we will prove another claim that helps us analyze Phase 2 of the two solutions.

\begin{claim}\label{9}
Let $OPT(n_1, n_2)$ denote the minimum amount of time needed to reach the goal $M$ from a state where you have 0 cookies, $n_1 \geq 0$ copies of item 1 have been purchased, and $n_2 \geq 0$ copies of item 2 have been purchased. If $n_1 + n_2 > 0$, then $OPT(n_1, n_2) < OPT(0, 0)$. The same statement holds if the final goal is not reaching $M$ cookies but to obtain $c_1 > n_1$ copies of item 1 and $c_2 > n_2$ copies of item 2.
\end{claim}
\begin{proof}
Consider the strategy $S(n_1, n_2)$ that mirrors the strategy of $OPT(0, 0)$ except that it doesn't purchase items when it has more items than $OPT(0,0)$. Specifically, when $OPT(0, 0)$ purchases its $n$th copy of item 1, $S$ will also choose to purchase the same item unless $n < n_1$, in which case $S$ will do nothing. When $OPT(0, 0)$ enters the Waiting Phase, $S$ will as well. Due to how $S$ is defined, when $S$ gets to the Waiting Phase, the generation rate of $S$ will be at least that of $OPT(0, 0)$. On the other hand, anything that happens before that will take $S$ less time than $OPT(0, 0)$ takes because $S$ starts with more items and this gives $S$ a higher generation rate and/or lets $S$ save time because $S$ may not have to purchase some items that $OPT(0, 0)$ does. Thus, it is clear that, $OPT(n_1, n_2) \leq S(n_1, n_2) < OPT(0, 0)$.
\end{proof}

The following claim is also necessary in conjunction with Claim~\ref{9}.
\begin{claim}\label{9_b}
If the optimal strategies ends up at the state $(c_1, c_2)$ just before Phase 3, the greedy algorithm will also reach the state $(c_1, c_2)$.
\end{claim}
\begin{proof}
In fact, a stronger claim is true: the greedy and optimal algorithms end up converging quite soon after Phase 1. To see why this is the case, consider a point in Phase 2 where the greedy algorithm has reached the state $(a, b)$, where both $ax_1$ and $bx_2$ are both larger than $q_2x_2$. This ensures that $(a, 0)$ and $(0, b)$ are also points in Phase 2. Suppose without loss of generality that it arrives at this state from the prior state $(a-1, b)$. Then, consider the optimal solution. At some point, the optimal solution must reach either $(c, b)$ where $c < a$ or $(a, d)$ where $d < b$. In the first case, we know that upon reaching $(c, b)$, the optimal solution will be following a greedy solution from that point forward. We also know that from the state $(a-1, b)$, the greedy solution chose to purchase item 1, indicating that from the state $(c, b)$, it is also more efficient to purchase item 1. This analysis holds true until the optimal solution reaches $(a, b)$, which means that the optimal solution has converged to the greedy solution. In the second case, we know that upon reaching $(a, d)$, the optimal solution will also be following a greedy solution from that point forward. We also know that the greedy solution must have purchased item 2 at some state $(e, d)$ where $e < a$ (because the greedy solution eventually reaches the state $(a-1, b)$). Thus, if it was more efficient to buy item 2 at the state $(e, d)$, the same will hold at the state $(a, d)$. This analysis applies until $d = b$, indicating that the optimal solution will also reach $(a,b)$.

Thus, the optimal and greedy solutions will definitely converge soon after Phase 1, and they will stay the same as long as Proposition 1 holds true, which is the entirety of Phase 2. Thus, both solutions will reach the same state $(c_1, c_2)$ at end of Phase 2.
\end{proof}

Finally, we need one last claim to analyze Phase 3 of the two solutions.  As described before, Phase 3 corresponds to the optimal solution only buying $w$ copies of the same item and taking $O_3$ time. 
\begin{claim}\label{10}
Let $T_3$ and $O_3$ denote the time taken by the greedy and the optimal solutions in this section, respectively. If the optimal solution only buys item 2 at the end, then $T_3 < O_3(\alpha_1(w\cdot \frac{x_2}{x_1} + 1))$. If the optimal solution only buys item 1 at the end, then $T_3 < O_3(\alpha_2(w+\frac{x_2}{x_1}))$.
\end{claim}
\begin{proof}
We'll begin by analyzing the case where the optimal solution buys $w$ copies of item 2 at the end. The greedy algorithm will buy $m$ copies of item $2$ and $n$ copies of item $1$. We know that if we look at the last item the greedy algorithm buys, the generation rate before buying that last item must be less than the optimal algorithm's end generation rate. Thus, either $mx_2 + (n-1)x_1 < wx_2$, so $n < (w-m)\frac{x_2}{x_1} + 1$, or $(m-1)x_2 + nx_1 < wx_2$, so $n < (w-m+1)\frac{x_2}{x_1}$. The two cases are similar, so we will just analyze the first one.

First, we observe that the greedy algorithm will never buy item $1$ if its cost is more than item $2$. Thus, at any point in time, item $1$ can cost at most $\alpha_1$ times as much as item $2$, or $y_1 < \alpha_1y_2$.

The total time needed by the greedy algorithm is the total time need to buy the $m$ copies of item $2$ plus the total cost of the $n$ copies of item $1$. Buying $m$ copies of item $2$ takes at most $\frac{m}{w} O_3$ (because the items get more expensive). Each copy of item $1$ can't take more than $\alpha_1$ times the maximum time needed to buy a copy of item $2$ in the greedy solution, which must be less than the maximum time need to buy item $2$ in the optimal solution, which must be less than $O_3$. Thus, buying $n$ copies of item $1$ can't take more than $n\alpha_1 O_3$. Thus, we have that $T_3 < O_3(\frac{m}{w} + n\alpha_1) < O_3(\frac{m}{w} + \alpha_1((w-m)\frac{x_2}{x_1}+1))$. This approximation ratio achieves its maximum when $m = 0$, giving $T_3 < O_3(\alpha_1(w\cdot \frac{x_2}{x_1} + 1))$. This completes the proof.

To prove the same result for the case where the optimal solution only buys item 1 at the end, we can observe that the greedy algorithm will never buy item 2 if $y_2 > \frac{x_2}{x_1} y_1$, so in the worst case, $y_2 < \alpha_2\frac{x_2}{x_1} y_1$ at any moment. The rest of the argument proceeds similarly.
\end{proof}

Finally, we provide a bound on the $w$ defined above. We will derive an expression for the case where the optimal solution only buys copies of item $2$; the bound for the other case can be derived similarly.
\begin{claim}\label{11}
$w < \frac{j\log(j\alpha_1)}{\log(\alpha_2)} + 1$, where $j = \big\lceil\frac{x_2}{x_1}\big\rceil$.
\end{claim}
\begin{proof}
If the optimal solution buys $k$ copies of item $2$ at the end, $k$ can not be too big because otherwise we could replace the purchase of the final $2$ with $j$ purchases of item $1$, which would give a greater rate increase and also take less time.

If the optimal solution buys $w$ copies of item $2$ at the end, it must have bought item $1$ right before that. Thus, at that moment, the optimal and greedy solutions match, and $y_1 < y_2$.
After $w-1$ purchase of item $2$, purchasing another copy of item $2$ would cost $y_2 \alpha_2^{w-1}$. On the other hand, purchasing $j$ copies of item $1$ would cost $\alpha_1y_1(1+\alpha_1+\dots+\alpha_1^{j-1}) < y_1j\alpha_1^j$. Technically, the generation rates when each of these items is purchased is different, but we can argue that the rates are all close enough to each other (say, within a small constant factor), and the overall argument still holds.

Then, it must be the case that
\begin{equation*}
y_1 \alpha_2^{w-1} < y_2\alpha_2^{w-1} < y_1j\alpha_1^j
\end{equation*}
Taking logs of both sides and rearranging gives the desired result.
\end{proof}

From \backref{Claim~\ref{8}}, \backref{Claim~\ref{9}}, \backref{Claim~\ref{9_b}}, \backref{Claim~\ref{10}}, and \backref{Claim~\ref{11}}, we can show that the greedy algorithm that uses the efficiency score is an approximation algorithm whose approximation ratio approaches 1 as $M \rightarrow \infty$.

Suppose we are given an instance of Cookie Clicker for 2 Items with sufficiently large $M$. Without loss of generality, suppose that $x_2 > x_1$. Let $Q_2$ be the larger positive solution to $x^2 + 2x = \frac{1}{\alpha_2-1}$.

Consider the greedy algorithm on this instance of Cookie Clicker. We will let $T(0, 0)$ be the amount of time it takes to reach the goal $M$. At the end of Phase 1, it will reach a rate of $G \geq Q_2x_2$. We denoted the amount of time taken in this phase by $T_1$. The amount of time it takes to reach this point is bounded above by the amount of time it takes to purchase $\lceil Q_2 \rceil$ copies of item $2$ while ignoring item 1, which is a function of the inputs $x_2,  y_2, \alpha_2$. Thus, $T_1 < d_1$, where $d_1 = f'(x_2, y_2, \alpha_2)$.

Next, suppose that when the greedy algorithm passes the rate $Q_2x_2$, it has $n_1$ copies of item 1 and $n_2$ copies of item 2. The greedy algorithm will continue to make locally optimal decisions from that point forward, which, as \backref{Claim~\ref{8}} shows, are globally optimal decisions. Let $T(n_1, n_2, c_1, c_2)$ denote the amount of time it takes the greedy solution to reach the end of the Phase 2. Now define $OPT(n_1, n_2, c_1, c_2)$ similarly for the optimal algorithm. \backref{Claim~\ref{9_b}} tells us that we can use the same values of $c_1$ and $c_2$. Then we know that $T(n_1, n_2, c_1, c_2) = OPT(n_1, n_2, c_1, c_2)$. Finally, we let $T_3$ denote the amount of time taken in Phase 3 of greedy algorithm. Then, using \backref{Claim~\ref{9}}, \backref{Claim~\ref{9_b}}, \backref{Claim~\ref{10}}, and \backref{Claim~\ref{11}}, we have that
\begin{align*}
T(0,0) &= T_1 + T_2 + T_3 \\
&= T_1 + T(n_1, n_2, c_1, c_2) + T_3 \\
&= T_1 + OPT(n_1, n_2, c_1, c_2) + T_3 \\
&< d_1 + OPT(0, 0, c_1, c_2) + d_2O_3,
\end{align*}
where $d_2$, depending on whether the optimal solution buys item 2 or item 1 at the end, has the form $\Big(\alpha_1\Big(\Big(\frac{\big\lceil\frac{x_2}{x_1}\big\rceil\log\big(\big\lceil\frac{x_2}{x_1}\big\rceil\alpha_1\big)}{\log(\alpha_2)} + 1\Big)\cdot \frac{x_2}{x_1} + 1\Big)\Big) = f^{*}(x_1, x_2, \alpha_1, \alpha_2)$.

Finally, note that in the optimal solution, the maximum possible generation rate that the solution will have before going into the Waiting Phase is $x_1\log_{\alpha_1}\frac{M}{y_1}+x_2\log_{\alpha_2}\frac{M}{y_2}$, which grows with $O(\log M)$. Thus, the Waiting Phase will take at least $\frac{M}{O(\log M)}$ time, so $OPT(0, 0) > c\cdot \frac{M}{\log M}$, where $c$ is a function of $(x_1, x_2, y_1, y_2, \alpha_1, \alpha_2)$ but is independent of $M$.

The approximation ratio of our greedy algorithm is $\frac{T(0, 0)}{OPT(0, 0)}$, and we know that
\begin{align*}
\frac{T(0, 0)}{OPT(0, 0)} &< \frac{d_1 + OPT(0, 0, c_1, c_2) + d_2O_3}{OPT(0, 0, c_1, c_2) + O_3} \\
&= \frac{d_1 + (d_2-1)O_3}{OPT(0,0)} + 1\\
&< \frac{d_1}{c} \cdot \frac{\log M}{M} + \frac{(d_2-1)O_3}{c}\cdot \frac{\log M}{M} + 1 \\
&= O\Big(\frac{\log M}{M}\Big) + O(\frac{O_3\log M}{M}) + 1.
\end{align*}

We can make $M$ large enough such that the generation rate $G$ after the Buying Phase is as big as we want it to be. If we have purchased $n_1$ copies of item 1 and $n_2$ copies of item 2, then $G_{n_1, n_2} = 1+n_1x_1+n_2x_2$. Using \backref{Lemma~\ref{4}}, the optimal algorithm will stop purchasing additional copies of item 2 when $M < y_2 \cdot \alpha_2^{n_2} (1 + \frac{G_{n_1, n_2}}{x_2})$. However, because it was worth it to purchase the $n_2$th copy, we know that
\begin{equation*}
M \geq y_2 \cdot \alpha_2^{n_2-1} \Big(1 + \frac{G_{n_1, n_2-1}}{x_2}\Big) > y_2 \cdot \alpha_2^{n_2-1} \cdot n_2 .
\end{equation*}
Thus $\log{M} = O(\log{n_2} + n_2) = O(n_2)$, so $n_2 = O(\log {M})$. Note that the cost of the last item purchased in this phase is $O(Mx_2/G_{n_1, n_2}) = O(M/n_2) = O(M/\log M)$. Because each of the $w$ items bought in Phase 3 can cost at most the cost of the last item, we have that the total cost of the last phase in the optimal solution is at most $w \cdot O(M/\log M)$. Recall from \backref{Claim~\ref{11}} that $w$ does not depend on $M$. The minimum rate in this phase is $G - wx_2 = O(\log M)$, so \backref{Lemma~\ref{2}} tells us that the total time cost is $O_3 = O (M/log^2M)$. Plugging this in for $O_3$ in the final expression, we get that the approximation ratio is
\begin{equation*}
O\Big(\frac{\log M}{M}\Big) + O\Big(\frac{1}{\log M}\Big) + 1 = 1 + O\Big(\frac{1}{\log M}\Big).
\end{equation*}
Thus, as $M \rightarrow \infty$, the $O(\frac{1}{\log M})$ term approaches $0$, and so the approximation ratio can be arbitrarily close to $1$ for sufficiently large $M$.

The above results can be extended to the case of $k$ items, as the local ``swapping" argument at the core of \backref{Claim~\ref{8}} works for any pair of consecutive item purchases. Thus, once the generation rate $G$ exceeds $\max_{i} q_ix_i$, where each $q_i$ is the smallest integer satisfying $q_i^2 + 2q_i \geq \frac{1}{\alpha_i-1}$, Phase 1 will end. Phase 1 could take longer in the $k$ item case than in the $2$ item case, but the amount of time it takes is still independent of $M$. \backref{Claim~\ref{9}} and \backref{Claim~\ref{9_b}} also hold for more items. Finally, \backref{Claim~\ref{10}} applies to any pair of items, so the number of items in Phase 3 is upper bounded by the value of $w$ derived from every pair of items, which is still a function of just the $x$'s and $\alpha$'s, independent of $M$. Thus, the same analysis applies, and we can achieve the same approximation ratio for larger $k$.
\end{proof}

For the rate-goal version of the problem, we believe that a similar approach could work, but it does not follow as easily. This is because the Waiting Phase does not exist, which is an essential part of proving the approximation ratio in the $M$ version. One approach that could work is to show that Phase 3 of the optimal and greedy solutions match exactly. If that can be proven, then we can prove an approximation ratio of $1+O\Big(\frac{R}{c^R}\Big)$.

\section{Negative Results}
\subsection{$R$ version is at least as hard as $M$ version}
We will begin by showing that the $R$ version of the problem is at least as hard as the $M$ version of the problem using a simple polynomial time reduction.
\begin{theorem}
The $M$ version of Cookie Clicker is polynomial time reducible to the $R$ version of Cookie Clicker
\end{theorem}
\begin{proof}
Suppose we are given an instance of the $M$ version of the problem with $k$ items having the parameters $(x_i, y_i, \alpha_i)$ for $1 \leq i \leq k$.

Then, we can construct an instance of the $R$ version of the problem with $k+1$ items having the parameters $(x_i, y_i, \alpha_i)$ for $1 \leq i \leq k$ and $(V, M, -)$ for item $k+1$, where $M$ is the target number of cookies in the simulated original problem, $V$ is the rate goal in the new problem, and - is an arbitrary value. We let $V$ be sufficiently large such that it is faster to purchase item $k+1$ to achieve the rate goal of $V$ than it is to achieve the same rate goal by purchasing only the first $k$ items.

Then, the optimal solution must purchase item $k+1$. To do so, we would need $M$ cookies as quickly as possible from the initial state by using just the first $k$ items. Thus, solving this specific instance of the $R$ version is equivalent to solving the $M$ version for $k$ items. Thus, the $M$ version is polynomial time reducible to the $R$ version and the $R$ version is at least as hard as the $M$ version.
\end{proof}
\subsection{Weak NP-hardness of $R$ version}
We will now prove weak NP-hardness of the $R$ version of $k$-item increasing-cost Cookie Clicker using a reduction from the weakly NP-hard problem \PARTITION\ \cite{Partition_Hardness1979}.

The problem \PARTITION\ is the following: Given a multiset $S$ of positive integers, can $S$ be partitioned into two subsets $S_1$ and $S_2$ such that the sum of the numbers in each subset is equal?

Because we want to prove NP-hardness results, we will use a decision version of the rate-goal Cookie Clicker problem rather than the original optimization version. The problem is the following: Given $0$ initial cookies, an initial cookie generation rate of $1$, and $k$ items described by tuples $(x_i, y_i, \alpha_i)$, is there a strategy that can obtain a rate of $R$ by target time $T$?

\begin{theorem}\label{rversionnphard}
The $R$ version of Cookie Clicker is weakly NP-hard.
\end{theorem}
\begin{proof}
Suppose we are given an instance of \PARTITION\ in the form of a set of positive integers $(a_1, a_2 \dots a_k)$ such that $\sum_{i=1}^{k} a_i = 2B$.
We will construct an instance of rate-goal Cookie Clicker such that solving this instance is equivalent to solving the input \PARTITION\ instance.

First, we choose $W = B^2+B+1$ and let $L$ represent an extremely large number such that it is never worth it to purchase two copies of any one item. Then, we construct the following instance of rate-goal Cookie Clicker:

\begin{itemize}
\item $(x_i, y_i, \alpha_i) = (a_i/W, a_i, L)$ for $1 \leq i \leq k$
\item $R = 1+B/W$
\item $T = B$
\end{itemize}

Our goal is to prove the following

\begin{lemma}\label{12}
A partition exists for the \PARTITION\ instance if and only if there exists a solution for the corresponding rate-goal Cookie Clicker instance which takes at most $B$ time.
\end{lemma}

\begin{proof}
First, note that a partition exists if and only if there exists a sequence of items whose rate gains add up to $B/W$.

We'll begin by assuming that a partition $a_{p_1}, a_{p_2}, \dots, a_{p_r}$ exists. Let $S_p$ denote the strategy that purchases the sequence of items $p_1, p_2, \dots, p_r$. We will show that this strategy $S_p$ takes at most $B$ time. 

Because item costs and rate gains are proportional in this instance of Cookie Clicker, the total cost (in cookies) of the items in strategy $S_p$ is $B$. Then, using \backref{Lemma~\ref{2}}, the total amount of time needed for strategy $S_p$ is
\begin{equation*}
B([p_1, \dots, p_r], 1) \leq \sum_{i=1}^{r}y_{p_i} = B
\end{equation*}
which proves the desired result.

Next, we will show that if no partition exists, then any solution to the Cookie Clicker instance will take more than $B$ time. If no partition exists, then no sequence of item purchases will add a rate gain of exactly $B/W$. Thus, any solution to Cookie Clicker must end at a rate of  $1+(B+n)/W$ for some positive integer $n$. Consider any  strategy $S$ that ends at such a rate. Again, because item costs and rate gains are proportional, the total cost (in cookies) of the items in strategy $S$ is $B+n$. Suppose that the items purchased in $S$ have indices $q_1, q_2, \dots, q_s$. Again, using \backref{Lemma~\ref{2}}, the total amount of time needed for strategy $S$ is
\begin{equation*}
B([q_1, \dots, q_s], 1) > \frac{\sum_{j=1}^{s}y_{q_j}}{1+(B+n)/W} = \frac{B+n}{1+(B+n)/W} = \frac{W(B+n)}{W+B+n}.
\end{equation*}
We want to show that
\begin{equation*}
\frac{W(B+n)}{W+B+n} > B
\end{equation*}
or equivalently, that
\begin{equation*}
W > \frac{B(B+n)}{n}
\end{equation*}
The final inequality is true as long as $W > B(B+1)$ because the right-hand side is maximized when $n=1$, so $W = B^2+B+1$ works. Thus, we have shown that if no partition exists, then there does not exist a solution to Cookie Clicker that takes at most $B$ time. This completes the proof of the lemma.
\end{proof}

\backref{Lemma~\ref{12}} is a proof that the reduction from \PARTITION\ holds, implying the statement of \backref{Theorem~\ref{rversionnphard}} that the R version of Cookie Clicker is weakly NP-hard. As of now, we have not been able to prove the same result for the M version of Cookie Clicker, so we provide a weaker hardness result for a variant of the M version of Cookie Clicker in the following section.
\end{proof}

\subsection{Cookie Clicker with Initial Cookies}
\label{Cookie Clicker with Initial Cookies}
We now focus on the more general version of the Cookie Clicker problem where you start with $z > 0$ initial cookies. Recall that previously, we only focused on the case where you start with $z = 0$ initial cookies. We will provide a pseudo-polynomial time algorithm for solving it and a weak NP-hardness proof. First, we list the inputs to this problem again, which are

\begin{itemize}
\item $z$, the initial number of cookies you start out with;
\item Vectors $X$, $Y$, and $\Alpha$, where each triple $(x_i, y_i, \alpha_i)$ represents the (generation rate gain, initial cost, cost gain) of each item. The vectors are of length $k$;
\item $r$, the initial generation rate; and
\item $M$, the target number of cookies.
\end{itemize}
The goal of this game is to find the optimal order of items to purchase to reach the goal $M$ as quickly as possible. This version of the game now uses the extra parameters $z$ and $r$, which were previously set to fixed values $z=0$ and $r=1$.

\subsubsection{Dynamic Programming Solution}

This section is not a negative result, but it describes a weakly-polynomial time solution to the Cookie Clicker with Initial Cookies problem.

The dynamic programming solution from \backref{Section~\ref{DP1}} can be modified slightly to solve this generalized problem. We will use the 2 Item case to illustrate our example.

Just as before, we know that the optimal solutions is to buy items whenever they are affordable or to enter the Waiting Phase Thus, if we have cookies left over and an item is affordable and beneficial, we will choose to buy that item right away. This means that right when the game starts, the strategy will be to buy some set of items all at once, until items are either not affordable or not beneficial, and then to wait until items become affordable again. This implies that each game state in the optimal solution can still be described by 2 numbers $(n_1, n_2)$, corresponding to the number of item 1 that has been purchased and the number of item 2 that has been purchased. It seems like the number of left over cookies would be a third parameter, but the number of left over cookies is determined entirely by $(n_1, n_2)$ based on this strategy.

Let $C_{n_1, n_2}$ be the cost of purchasing $n_1$ copies of item 1 and $n_2$ copies of item 2 at the very beginning of the game. If $C_{n_1, n_2} < z$, the number of left over cookies is just $k - C_{n_1, n_2}$, and if $C_{n_1, n_2} > z$ the number of left over cookies is exactly $0$. Let $L_{n_1, n_2} = \max\Big(k - C_{n_1, n_2}, 0\Big)$ denote the number of leftover cookies.

 Then, since each state can be described by $2$ numbers, we can derive the recurrence:
\begin{align*}
DP[a][b] = \min\Big(&\frac{M-L_{a,b}}{g_{ab}}, \\
&\frac{\max(y_1\cdot\alpha_1^a-L_{a,b}, 0)}{g_{ab}} + DP[a+1, b], \\
&\frac{\max(y_2\cdot\alpha_2^b-L_{a,b}, 0)}{g_{ab}} + DP[a, b+1]\Big)
\end{align*}
Here, $g_{ab} = r + ax_1 + bx_2$.

We can generalize this formula to $k$ items and it will still hold. Thus, dynamic programming provides a weakly-polynomial time solution to the Cookie Clicker with Initial Cookies problem.
\subsubsection{Weak NP-hardness of Cookie Clicker with Initial Cookies}

We will now prove that the Cookie Clicker with Initial Cookies problem is NP-hard.

\begin{theorem}\label{13}
Cookie Clicker with Initial Cookies is NP-hard.
\end{theorem}
\begin{proof}
As before, our strategy will be to use a reduction from \PARTITION. Suppose we are given an instance of \PARTITION\ in the form of a set of positive integers $(a_1, a_2 \dots a_k)$ such that $\sum_{i=1}^{k} a_i = 2B$.
We will construct an instance of Cookie Clicker with Initial Cookies such that solving Cookie Clicker with Initial Cookies will solve \PARTITION.

Let $A$ be some big number (for example, $A = 1000B$), and let $L$ be an extremely large number such that an optimal strategy should only buy at most 1 of each item. We can construct an instance of Cookie Clicker with Initial Cookies with the inputs set as follows:
\begin{itemize}
\item $z = kA+B$
\item $(x_i, y_i, \alpha_i) = (a_i+A, a_i+A, L)$ for $1 \leq i \leq k$
\item $(x_i, y_i, \alpha_i) = (A, A, L)$ for $n+1 \leq i \leq 2k$
\item $r = 0$
\item $M = kA+B+1$
\end{itemize}

We will now prove the following lemma.
\begin{lemma}\label{14}
A partition exists for the \PARTITION\ instance if and only if there exists a solution to the corresponding Cookie Clicker with Initial Cookies instance which takes at most $\frac{M}{kA+B}$ time.
\end{lemma}

\begin{proof}
First, suppose that a \PARTITION\ solution exists. That means we can choose some set of the integers $(a_1, a_2 \dots a_k)$ such that they sum to $B$. Equivalently, this means we can use our initial $z = kA+B$ cookies to buy $k$ total items for a price of $kA+B$ at the very beginning of the game. We then wait until we have $M$ cookies. The total amount of time it takes to reach $M$ using this strategy is $\frac{M}{kA+B}$, which proves the first half of the lemma.

Now, suppose that there exists a solution to the Cookie Clicker with Initial Cookies instance which takes at most $\frac{M}{kA+B}$ time. Recall again the optimal strategy has a Buying Phase and a Waiting Phase. We will analyze what rate the strategy ends up with at the end of the Buying Phase with the goal of showing that a solution that takes at most $\frac{M}{kA+B}$ time must end at a final rate of $kA+B$. We will do so via contradiction.

If the strategy doesn't end up at a rate of $kA+B$, there are two possible cases. We will assume that each case is true and derive a contradiction.

\textbf{Case 1:} The strategy ends at a rate less than $kA+B$

If the strategy ends at a rate of $kA+B-j$ for $j > 0$, then it must have spent $kA+B-j$ purchasing items at $t=0$ and then waited from that point forward. The total time that this strategy takes is $\frac{M-j}{kA+B-j}$, which we claim is always greater than $\frac{M}{kA+B}$. It's easy enough to verify that
\begin{equation*}
\frac{M}{kA+B} < \frac{M-j}{kA+B-j}
\end{equation*}
or equivalently
\begin{equation*}
-jM < -j (kA+B)
\end{equation*}
which follows from $kA+B < M$. Thus, we have a contradiction, and this case is impossible.

\textbf{Case 2:} The strategy ends at a rate greater than $kA+B$.

First, note that after buying any $k$ items (or particular combinations of $k-1$ items) at $t=0$, there will not be enough cookies left over to purchase the next item right away. The Cookie Clicker with Initial Cookies problem then reduces to the original Cookie Clicker problem where you have no cookies at the start. We can then recall from \backref{Corollary~\ref{5}} that if you have generation rate $G$, it is only worth it to buy an item with rate increase $x$ and cost $y$ if
\begin{equation*}
G \leq \frac{Mx}{y}-x.
\end{equation*}
Because $x_i = y_i$ for all $i$, this becomes
\begin{equation}
G \leq M-x_i.
\end{equation}

In the case where a solution buys $k$ items at $t=0$, $G \geq kA$. Then $G + x_i \geq G + A \geq (k+1) A > M$, so it is no longer worth it to buy any items after purchasing $k$ items. However, the $k$ items cannot have total cost greater than the initial amount of cookies, which is $kA+B$, and we know that their total cost is not exactly $kA+B$, so the final rate $G$ is at most $kA+B-1$. This does not match the assumption in this case that the final rate is greater than $kA+B$.

Now, consider the case where the solution buys $k-1$ items at $t=0$. Then equation (5) tells us that item $i$ is only worth purchasing if $M \geq G + x_i$. Thus, the final rate after purchasing the $k$th item, $G+x_i$, is upper bounded by $M = kA+B+1$. Because we are currently considering the case where the final rate is greater than $kA+B$, the only possible final rate for this case is then $kA+B+1$.

Thus, we have reduced this case to the specific scenario where $k-1$ items are purchased at time $t=0$, giving a rate of $(k-1)A + c$ for some $ 0 \leq c \leq 2B$. Then, the strategy purchases another item as soon as it can, ending up at a rate of $kA+B+1$. After purchasing the first $k-1$ items, the strategy will have $A+B-c$ cookies left. The cost of the last item, which will push the rate up to $kA+B+1$, will be $kA+B+1 - (k-1)A - c = A+B-c+1$. Thus, this strategy needs exactly 1 more cookie to purchase this last item.

The amount of time this strategy takes is
\begin{equation}
\frac{1}{(k-1)A+c} +\frac{M}{kA+B+1} \geq \frac{1}{(k-1)A+2B} +\frac{M}{kA+B+1} .
\end{equation}
We want to check that the quantity on the right-hand side of (6) is greater than $\frac{M}{kA+B}$. Indeed,
\begin{align*}
    && \frac{1}{(k-1)A+2B} +\frac{M}{kA+B+1}  &>  \frac{M}{kA+B} \\
\iff&& \frac{1}{(k-1)A+2B} &>  \frac{M}{(kA+B)(kA+B+1)} \\
\iff&& \frac{1}{(k-1)A+2B} &>  \frac{1}{kA+B} \\
\iff&& kA+B &>  (k-1)A+2B \\
\iff&& A &> B,
\end{align*}
which is true. Thus, any strategy that ends at a rate greater than $kA+B$ will also take more than $\frac{M}{kA+B}$ time. Again, we have a contradiction, and this case is impossible.

Therefore, if a solution exists that takes at most $\frac{M}{kA+B}$ time, it must end at a rate of $kA+B$. Then, since the cost and rate gains are the same for each item, the solution must have purchased a set of items that have a total cost (in cookies) of $kA+B$. Then, this set of items corresponds to a set of $a_i$'s that sum to $B$, which means that a solution to the \PARTITION\ instance exists. Thus, we have shown that if a solution takes at most $\frac{M}{kA+B}$ time, then a partition exists for the original \PARTITION\ instance. This completes the proof of \backref{Lemma~\ref{14}}.
\end{proof}

We have shown that given an instance of \PARTITION, which is NP-complete, we can construct an instance of Cookie Clicker with Initial Cookies in polynomial time such that being able to solve Cookie Clicker with Initial Cookies means being able to solve the instance of \PARTITION. Thus, \backref{Theorem~\ref{13}} is proven and Cookie Clicker with Initial Cookies itself is NP-hard.
\end{proof}

As a note, this result was proven for the $M$ version of Cookie Clicker with Initial Cookies, so it must hold for the $R$ version too because the $R$ version is at least as hard as the $M$ version.

\subsection{Cookie Clicker with Discrete Timesteps is Strongly NP-hard}
\label{Discrete Cookie Clicker is Strongly NP-hard}

Another variant of Cookie Clicker is Cookie Clicker with discrete timesteps. In all previous versions, we have been analyzing the game in continuous time. In this model, we can think of the generation rate as an ``income'' instead, where you receive your income (some number of cookies) after every discrete time step. This model can be shown to be NP-hard via a reduction from the strongly NP-hard problem 3-\PARTITION\ \cite{Partition_Hardness1979}.

The Cookie Clicker with Discrete Timesteps problem can be formally stated as the following: Given $0$ initial cookies, an initial income $r$, $n$ items described by tuples $(x_i, y_i, \alpha_i)$, and the rule that you receive your income after every timestep, is there a strategy that can obtain $M$ cookies by target time $T$?

The problem 3-\PARTITION\ is the following: Given a multiset $S$ of $k=3m$ integers, can $S$ be partitioned into triplets $S_1, S_2, \dots, S_m$ such that the sum of the numbers in each subset is equal?

\begin{theorem}
Cookie Clicker with Discrete Timesteps is strongly NP-hard
\end{theorem}
\begin{proof}
We reduce from 3-\PARTITION. Suppose we are given an instance of 3-\PARTITION\ $(a_1, a_2, \dots, a_k)$, such that $\sum_{i=1}^{k} a_i = A$.

We will encode 3-\PARTITION\ as Cookie Clicker with Discrete Timesteps as follows. Choose a number $B > \frac{Ak}{3}$. As before, let $L$ be a large enough number such that it is never worth it to buy two of any one particular item. We then construct the following Cookie Clicker with Discrete Timesteps instance:

\begin{enumerate}
\item \ $(x_i, y_i, \alpha_i) = (a_i, B\cdot a_i, L)$.
\item \ $M$ = $\frac{A}{2}\big(\frac{k}{3}-1\big) + 2B \cdot \big(\frac{3BA}{k} + A\big)$.
\item \ $r = \frac{3BA}{k}$.
\item \ $T = \frac{k}{3} + 2B$
\end{enumerate}

We will prove that there exists a solution to the 3-\PARTITION\ instance if and only if there exists a solution to the Cookie Clicker instance which reaches $M$ in time $\frac{k}{3} + 2B$ or less. Recall that the optimal solution must proceed in two distinct phases: the Buying Phase and the Waiting Phase.

First, note that after time step $\frac{k}{3}$, ignoring any extra income we get from buying items in those timesteps, we will have produced $BA$ cookies just from our initial generation rate. Thus, it will be possible to purchase every single item by the end of time step $\frac{k}{3}$. Since it is always better to purchase items earlier rather than later, this means that the Buying Phase will last at most $\frac{k}{3}$ turns.

Next, note that buying every single item results in a final generation rate of $\frac{3BA}{k} + A$. Thus, the generation rate is always upper bounded by $\frac{3BA}{k} + A$. At the end of time step $\frac{k}{3}$, the total amount of cookies generated will be at most $BA + \frac{Ak}{3}$. Because $\frac{Ak}{3} < B$ and all item costs are multiples of $B$ (and thus at least $B$), we know that any cookies generated from our items and not from the original income will not increase our ability to purchase items in the Buying Phase. In other words, the cookies that our items generate will not improve our buying power, and we essentially get $\frac{3BA}{k}$ to spend every turn. We can also conclude that the Buying Phase will last exactly $\frac{k}{3}$ turns.

Next, note that because the cost of every item scales linearly with the increase in generation rate, spending cookies on items will always produce the same increase in generation rate per cookie. Thus, the best way to spend cookies is to spend them as early as possible, because this maximizes the amount of time that the increase in generation rate is present.

Finally, $T$ is large enough that it is always worth it to purchase the first copy of each item rather than foregoing the purchase. Thus, the optimal strategy is to spend as many of your cookies as possible at every time step in the Buying Phase, and then to enter the Waiting Phase. The fastest way to reach $M$ will be to spend all $\frac{3BA}{k}$ of your generated cookies on every time step in the Buying Phase. This is only possible if the numbers $(a_1, a_2, \dots, a_k)$ can be partitioned into subsets such that for each subset, the total sum is $\frac{3A}{k}$, which is exactly the 3-\PARTITION\ problem. The maximum attainable value of $M$ assuming a 3-\PARTITION\ exists is the value we chose for $M$ in the reduction.
\end{proof}
Once again, note that this result also holds for the corresponding $R$ version of the Cookie Clicker with Discrete Timesteps problem.

\section{Conclusion}

Cookie Clicker, while a seemingly simple game, gives rise to many interesting optimization problems. We analyzed these problems through the context of dynamic programming, approximation algorithms, and NP-hardness. For specific variants of Cookie Clicker, we classified the structure of optimal solutions, thereby limiting our search space for such solutions. This allowed us to devise polynomial time algorithms for solving the problem. For more general variants of Cookie Clicker, we proved NP-hardness results via reductions from \PARTITION\ and 3-\PARTITION. Although these problems are NP-hard, their solutions can be approximated very well with a greedy algorithm based on a specific efficiency metric, and we can prove an approximation ratio guarantee that approaches $1$ when the input parameter approaches infinity.


Here are a few conjectures which experimentally appear to be true, but have not been proved.
\begin{itemize}
\item \textbf{Fixed-Cost Case for $k$ items:} \emph{In any solution, if there are two items such that item 1 is cheaper and has a lower rate increase to cost ratio than item 2, then item 1 will never be bought after item 2.} This conjecture would imply that any solution to the fixed-cost case is in ``sorted" order, where the cheaper and less efficient items come first. This matches current experimental results. If this conjecture is true, then it would lead to a polynomial-time solution for small values of $k$ that involves finding the points of transition between buying one item as opposed to another. Analyzing a specific subset of local optimizations different from the ones analyzed in this paper could lead to insight on this conjecture.
\item \textbf{Increasing-Cost Case for $k$ items:} \emph{The problem of minimizing the amount of time needed to reach $M$ cookies from a starting state of $0$ initial cookies, an initial generation rate of $1$, and a set of $k$ items whose costs increase exponentially is weakly NP-hard.} This conjecture would be interesting because we would then have a very simple approximation algorithm whose approximation ratio approaches $1$ for sufficiently large $M$ for an NP-hard problem.
\end{itemize}
Other interesting directions to explore include the following:
\begin{itemize}
\item \textbf{Different Cost Increase Dynamics:} Instead of having item costs increase exponentially, have item costs increase additively or in some manner that matches economic situations more closely.
\item Relate the incremental game model to more real-world situations.
\end{itemize}

\section{Acknowledgements}
The authors would like to acknowledge the support of CREST, JST, Grant No. JPMJCR1402 and KAKENHI, JSPS, Grant No. 15K11985.

\let\realbibitem=\bibitem
\def\bibitem{\par \vspace{-1.2ex}\realbibitem}

\bibliography{CookieClicker}
\bibliographystyle{alpha}

\end{document}